\documentstyle[aps,twocolumn,prb]{revtex}
\begin{document}
\draft
\title{Theoretical study of O adlayers on Ru\,(0001)}
\author{C. Stampfl and M. Scheffler}
\address{
Fritz-Haber-Institut der Max-Planck-Gesellschaft \\
Faradayweg 4-6, D-14\,195 Berlin-Dahlem, Germany
}
\maketitle

\begin{abstract}
Recent experiments performed at high pressures indicate that ruthenium
can support unusually high concentrations of oxygen at the surface.
To investigate the structure and stability of high coverage oxygen
structures, we performed
density functional theory calculations, within the generalized
gradient approximation, for
O adlayers on Ru\,(0001) from low coverage up to a full monolayer.
We achieve quantitative agreement with previous low energy
electron diffraction intensity analyses for the
$(2 \times 2)$ and $(2 \times 1)$ phases
and predict that an O adlayer with a $(1 \times 1)$ periodicity
and coverage $\Theta$=1
can form on Ru\,(0001), where the O adatoms occupy hcp-hollow sites.
\end{abstract}

\section{Introduction}

The interaction of oxygen with metal surfaces forms the
basis of many
technologically important processes, for example,
bulk oxidation, corrosion, and heterogeneous catalysis.
It is therefore of great interest to obtain a detailed understanding
of the changes in the  atomic
and electronic structure  that oxygen adsorption
often induces due to the formation of strong chemical bonds
\cite{starke,besenbacher}.
The behavior of O on metal surfaces is quite varied and depends
markedly on the coverage and temperature, and on the orientation
of the surface of the particular metal.
Generally, the close-packed surfaces are more stable
against reconstruction; often, however,
significant atomic relaxations of the substrate
are induced by O adsorption \cite{starke}.
On Zr, the O atoms apparently form an underlayer
\cite{hui} even for the close-packed (0001) surface.
At higher coverages of oxygen, at elevated
temperatures, oxide-like structures can form
on a number of metal surfaces \cite{besenbacher}.

 From recent experiments of the catalytic
oxidation of carbon monoxide, performed at high pressures, there is
evidence that Ru\,(0001) can support unusually high concentrations
of oxygen at the surface \cite{peden2,peden}. In order to investigate
the structure and stability of high coverage oxygen structures, we
performed extensive density functional theory
calculations for various O adlayers on Ru\,(0001). In particular,
for the two ordered phases, $(2 \times 2)$ \cite{lindroos} and
$(2 \times 1)$ \cite{pfnur},
which form at room temperature
under ultra high vacuum (UHV) conditions
for coverages $\Theta=1/4$ and  $\Theta=1/2$, respectively,
as well as for an artificial $(2 \times 2)$-3O adlayer with
coverage $\Theta=3/4$, and
for several higher coverage
$(1 \times 1)$-O structures with $\Theta=1$.
Here, $\Theta$ is defined to be the
ratio of the number of adsorbate
atoms to the number of atoms in an ideal substrate layer.
The calculations reveal that although a $(1 \times 1)$ phase is not
observed to form under UHV conditions using molecular oxygen, perhaps
due to the presence of activation energy barriers for dissociation
of O$_{2}$, the
adsorption of O in a $(1 \times 1)$ adlayer structure with coverage
$\Theta=1$ is exothermic and should be able to form when energy
barriers can be overcome.

\section{Calculation method}

We use density functional theory (DFT) and employ
the generalized gradient approximation (GGA) of
Perdew {\em et al.} \cite{perdew} for the exchange-correlation functional.
The surface is modelled using the supercell approach and we employ
{\em ab initio}, fully separable
pseudopotentials created by the scheme of Troullier and Martins
\cite{troullier}, and a plane-wave basis set for expansion
of the electronic wave functions.
The calculations are performed using a $(2 \times 2)$ surface
unit cell and four layers of Ru\,(0001)  with a vacuum region corresponding
to thirteen such layers. The  O atoms
are adsorbed  on one side of the slab \cite{neugebauer}.
The energy cut-off is taken to be 40 Ry
with three special {\bf k}-points in the surface
 Brillouin zone \cite{cunningham}.
The calculation scheme \cite{stumpf} affords
simultaneous relaxation of the electrons and atoms using
damped Newton dynamics.
We relax  the positions of the O atoms and the Ru atoms in the top two
layers,
keeping the lower two Ru layers fixed.
For the higher coverage $(1 \times 1)$-O structures
we use a larger energy cut-off of 60 Ry
and  14  {\bf k}-points in the irreducible part of the surface
Brillouin zone of a $(1 \times 1)$ surface unit cell.

\section{O on Ru\,(0001)}
\subsection{$(2 \times 2)$-O/Ru\,(0001)}

We performed calculations for O in the fcc-hollow site
(no atom in the layer beneath the site) and in
the hcp-hollow  site.
 From our calculations we find that
the hcp-hollow site is energetically clearly favorable.
This is in agreement with the site determined by a dynamical LEED
intensity  analysis \cite{lindroos}.
The binding energy of O (relative to a free O atom)
 in the hcp-hollow site is 5.55 eV
and in the fcc-hollow site it is 5.12 eV  (see Tab.~I).
We can compare this result with calculations for
a single O adatom on an eighteen atom cluster model
representing Ru\,(0001) \cite{anderson}, using an atom superposition
and electron delocalization molecular orbital ASED-MO method. In this study
it was found that  the {\em fcc-hollow} site is
the most favorable with a binding energy of 5.6 eV.
The hcp-hollow site was the next most favorable with a
binding energy of 5.3 eV.
Such cluster geometries are often regarded as
representing the situation of an
{\em isolated} adsorbate on an extended surface.
Our calculated coverage dependence of
the adsorption site of O, discussed below (see Fig. 4a),
however, suggests that an isolated O adatom will occupy the
hcp-hollow site.

The atomic geometry of $(2 \times 2)$-O/Ru\,(0001) is
displayed in Fig. 1.
In Tab. II we compare the theoretically obtained structural
parameters
with those obtained by the LEED analysis \cite{lindroos}.
For ease of comparison, we use the same notation
as in the LEED study.
We note  that the theoretical
lattice constant is 1.9 \% larger than the experimental one.
Rather than giving a detailed comparison of all the
substrate relaxations, we simply refer to Tab. II, from which the high level
of agreement with the LEED analysis can immediately be assessed.
The theoretically obtained O-Ru  bond length of 2.08 \AA\, is
 slightly longer than
the LEED-determined value of 2.03 \AA\,.
The first Ru-Ru interlayer spacing is found to be contracted by
2.7 \%
with respect to the bulk value
(using the centers of gravity of the first and second buckled Ru layers).
This agrees well with the value
determined from the LEED analysis of 2.1 \%.
Such a contraction is
in contrast to many similar systems
formed by adsorption of strongly electronegative species,
where instead there is often an {\em expansion} induced by the
adsorbate \cite{starke}.
We find the
top Ru-Ru interlayer spacing of Ru\,(0001) to be
contracted by 2.5 ~\% which is close to the LEED-determined
value of  2.3~\% \cite{over}.
The  contraction  of the top interlayer spacing of the  clean
surface is therefore not removed by oxygen adsorption at 1/4 monolayer.

\subsection{$(2 \protect\times 1)$-O/Ru\,(0001)}

At higher oxygen coverage, namely, $\Theta=1/2$,
a  $(2 \times 2)$ LEED pattern is observed which corresponds to
three rotated domains, each of $(2 \times 1)$  periodicity \cite{pfnur}.
We performed calculations for O in the fcc- and hcp-hollow sites.
The hcp-hollow site is again found to be energetically the most favorable
with a binding energy of  5.28 eV; that for
the fcc-hollow site is 5.00 eV (see Tab.~I).
The theoretical identification of the hcp-hollow site for O in the
$(2 \times 1)$  structure
is in accord with the LEED determination for the adsorption site
\cite{pfnur}.

The atomic structure of $(2 \times 1)$-O/Ru\,(0001)
is depicted in Fig.~2.
The O atoms adsorb in ``off'' hcp-hollow sites, i.e., they are displaced
from the center of the hcp-hollow site. In addition,
complex O induced relaxations of the substrate
occur,  including ``row-pairing'' and
buckling of the substrate layers.
The determined O-Ru bond length, and the lateral and vertical
relaxations are given in Tab. III where they are compared with the
results obtained from the LEED analysis \cite{pfnur}.
Again, it can quickly be seen that quantitative agreement is achieved.
We do note one deviation however:
The directions of the lateral displacement of atom D,
$\Delta d_{\parallel}$(D), have the opposite
signs. That is,  we
obtain row-pairing of the Ru atoms in the {\em second} Ru layer, as well as
in the first layer,
and the LEED analysis does not.
We found that relaxing the third Ru layer does not change this result.
The O-Ru bond length of 2.07~\AA\,
is very similar to that which  we determined for the lower coverage
$(2 \times 2)$ structure which was 2.08 \AA\,.
The value is again somewhat larger than that of 2.02~\AA\, as obtained from
the LEED analysis.
The first two Ru-Ru interlayer spacings,
defined with respect to the
centers of gravity of the buckled atomic layers, correspond to the
bulk value to within 0.01~\AA\,, for both the DFT-GGA and LEED results.

\subsection{$(1 \times 1)$-O/Ru\,(0001)}

We now investigate the structure and stability of $(1 \times 1)$-O
adlayers with coverage  $\Theta=1$. We performed calculations for
O in the on-top, bridge, fcc-, and hcp-hollow sites.
The obtained binding energies are collected in Tab. I.
The hcp-hollow site is energetically preferred with a binding
energy of 4.87 eV. The
fcc-hollow site is the next most favorable having a binding
energy of 4.81 eV.
Table I also lists the binding energy differences for the
adsorption sites tested, with respect  to the binding energy of O in the
respective hcp-hollow sites.
The atomic geometry of
$(1 \times 1)$-O/Ru\,(0001) is shown in Fig.~3
and the O-Ru bond length and structural parameters are given in Tab. IV.
It is noticeable that the O-Ru bond length
of 2.03 \AA\, is slightly shorter than that of the lower coverage
structures.
The first Ru-Ru interlayer spacing is found to be
{\em expanded} by 2.7  \%.

For comparison we also performed
calculations for O in the hcp-hollow site
using an energy cut-off of 40 Ry
and three special {\bf k}-points in the Brillouin zone \cite{cunningham}.
As can be seen from Tab.~IV, the resulting binding energy is
4.84 eV, i.e., only 0.03 eV less than the result of the  more
accurate calculation, and the resulting structural parameters differ by
at most 0.02~\AA\,.

The value of the binding energy
of O in the hcp-hollow site on Ru\,(0001) at coverage $\Theta=1$
shows that the adsorption is exothermic and
indicates that the $(1 \times 1)$ adlayer structure should be
able to form. That
is, the binding energy is larger (by $\approx$ 1.8 eV per atom)
 than that which the O atoms have in O$_{2}$.
Under UHV conditions, however, the $(2 \times 1)$ phase
is the terminal one.
The reason that the $(1 \times 1 )$ structure
does not form under UHV conditions
could be due to a kinetic hindering of the
dissociation of O$_{2}$ induced by the $(2 \times 1)$ structure.
Interestingly, on a {\em stepped} Ru\,(0001) surface, the formation of a
$(1 \times 1 )$ structure for coverage $\Theta=1$ has been reported,
which is stable to 600 K  \cite{parott}.
On the stepped surface, it is possible that
step edges may act as sites over which
dissociation of O$_{2}$ can occur.
If atomic, as opposed to molecular, oxygen would be used then it is
probable that the $(1 \times 1 )$ phase will also be observed
on Ru\,(0001) under UHV.
This result could have implications for heterogeneous catalytic
reactions in which dissociative adsorption of O$_{2}$ is a necessary
reaction step (often rate-limiting) in that if atomic oxygen would
be used the kinetics may be greatly altered;
it also raises the question if other metal surfaces can also
support high O coverages.

\section{Coverage dependence}

As we have seen from above,
the hcp-hollow site is the preferred adsorption site for O on Ru\,(0001)
at all the coverages investigated.
This is consistent with the common
trend that atoms strongly chemisorbed on transition metal surfaces
usually occupy the site
that the next substrate layer would occupy.

In Fig. 4a the binding energy
of O in the fcc- and hcp-hollow sites as a function of coverage is
displayed.
With increasing coverage
the binding energy becomes less  favorable which reflects a repulsive
interaction between the adsorbates and implies
that no island formation is expected to occur in the the coverage regime
of $\Theta=1/4$ to $\Theta=1$.
Figure 4a also shows that the difference in binding energy
between the fcc- and hcp-hollow sites becomes less, as the
coverage increases.
As noted above, we also performed calculations for a
structure with coverage $\Theta=3/4$.
In this structure
we placed O atoms in hcp-hollow sites in the $(2 \times 2)$ surface unit
cell.
The O-Ru bond length is 2.07 \AA\, which is similar to that of the two
lower coverage structures, and the first
Ru-Ru interlayer spacing is expanded by 1.8 \%.
The corresponding work function change as a function of coverage
is shown in Fig.~4b.
The experimental results of Surnev {\em et al.}
\cite{surnev} are included
for comparison where good agreement between theory and experiment is
obtained.
The large increase in work function
reflects the strong electronegative nature of O.

\section{Conclusion}

{}From our calculations we achieve quantitative agreement with previous
structural determinations for $(2 \times 2)$-O and $(2 \times 1)$-O/Ru\,(0001)
structures, with respect to the preferred adsorption site
and the detailed atomic positions in the surface region.
One minor difference is found for the $(2 \times 1)$ structure
which would be worth a re-analysis of the LEED data.
We  predict that an O adlayer with a $(1 \times 1)$ periodicity
and coverage $\Theta$=1
should be able to form on Ru\,(0001), where the O atoms occupy
hcp-hollow sites.
For the coverage regime $\Theta$=1/4 to  $\Theta$=1 there is
no indication of island formation.
The corresponding work function change agrees well with experiment.

\vspace{0.7cm}
{\bf Acknowledgements}
\vspace{0.5cm}

We wish to thank Martin Fuchs for his
invaluable help in creating the pseudopotentials.

%----------------------- FIGURE CAPTIONS ---------------------

\begin{figure}
\caption{Top view (a) and side view (b) of the atomic
 geometry of $(2 \times 2)$-O/Ru\,(0001).
The arrows indicate the direction of the displacements of the substrate atoms
with respect to the bulk terminated positions.
The dashed line  in (a) indicates the plane of the cross-section used
in (b).  Small dark grey circles represent
oxygen atoms and large grey circles represent Ru atoms.
Interlayer spacings are given in \AA ngstrom.}
\end{figure}

\begin{figure}
\caption{Top view (a) and side view (b) of the atomic
 geometry of $(2 \times 1)$-O/Ru\,(0001).
The arrows indicate the direction of the atomic displacements.
The dashed line in (a) indicates the plane of the cross-section used
in (b).
 Small dark grey circles represent
oxygen atoms and large grey circles represent Ru atoms.
Interlayer spacings are given in \AA ngstrom.}
\end{figure}

\begin{figure}
\caption{Top view (a) and side view (b) of the atomic
 geometry of $(1 \times 1)$-O/Ru\,(0001) with O in the hcp-hollow site.
The arrows indicate the direction
of the displacements of the substrate atoms
with respect to the bulk terminated positions.
 Small dark grey circles represent
oxygen atoms and large grey circles represent Ru atoms.
Interlayer spacings are given in \AA ngstrom.}

\end{figure}

\begin{figure}
\caption{(a) Binding energy of O on Ru\,(0001) for O in the
fcc- (dashed line) and in the hcp-hollow site
(continuous line). (b) Work function
change as a function of coverage, $\Theta$.
The experimental results, shown as open circles,
 are from Ref.~\protect\cite{surnev}.}
\end{figure}

%----------------------- TABLES ---------------------------------------------
\onecolumn
\newpage
\begin{table}
\begin{tabular}{l|l}
Sites tested   & \hspace{0.8cm}on-top  \hspace{1.0cm} bridge \hspace{0.3cm}
  fcc-hollow  \hspace{0.3cm}  hcp-hollow  \\
\hline
 $(2 \times 2)$ & \hspace{0.8cm} \hspace{1.0cm}      \hspace{3.2cm}
5.12 \hspace{1.0cm} 5.55 \\
 $\Delta\,E$  & \hspace{0.8cm}       \hspace{1.0cm}      \hspace{3.2cm} 0.43
 \hspace{1.0cm} 0.00 \\
\hline
 $(2 \times 1)$ & \hspace{0.8cm}       \hspace{1.0cm}    \hspace{3.2cm} 5.00
 \hspace{1.0cm} 5.28 \\
 $\Delta\,E$  & \hspace{0.8cm}       \hspace{1.0cm}      \hspace{3.2cm} 0.28
 \hspace{1.0cm} 0.00 \\
\hline
 $(1 \times 1)$ & \hspace{0.8cm}  3.62 \hspace{1.0cm} 3.93 \hspace{1.6cm} 4.81
 \hspace{1.0cm} 4.87 \\
 $\Delta\,E$ & \hspace{0.8cm}  1.25 \hspace{1.0cm} 0.94 \hspace{1.6cm} 0.06
 \hspace{1.0cm} 0.00 \\
\end{tabular}
\caption{Binding energies (in eV) of O on Ru\,(0001) relative to the
free O atom for the surface  structures investigated.
The binding energy differences, $\Delta\,E$,
defined relative to the value for the respective  hcp-hollow sites,
are also given.}
\end{table}

\begin{table}
\begin{tabular}{l|c}
 &  $(2 \times 2)$-O/Ru\,(0001)  \\
\hline
Structural parameters (\AA)  & \hspace{0.6cm}O-Ru \hspace{1.0cm}
  $\Delta d_{\parallel}$(A) \hspace{0.3cm}
 $\Delta d_{\parallel}$(D) \hspace{0.3cm}  $\Delta d_{\rm z}$ (A)
\hspace{0.4cm} $\Delta d_{\rm z}$ (B) \hspace{0.5cm}
$\Delta d_{\rm z}$ (C)  \\
\hline
LEED  & \hspace{0.2cm}  2.03 \hspace{1.5cm}
  0.09 \hspace{1.0cm} 0.01
\hspace{0.5cm} $-$0.05 \hspace{0.7cm} $-$0.12 \hspace{1.0cm} $-$0.08  \\
DFT-GGA  & \hspace{0.2cm}  2.08 \hspace{1.5cm}
  0.07 \hspace{1.0cm} 0.01
 \hspace{0.5cm} $-$0.04  \hspace{0.7cm} $-$0.11  \hspace{1.0cm} $-$0.03 \\
\end{tabular}
\caption{Structural parameters for $(2 \times 2)$-O/Ru\,(0001)
with O in the hcp-hollow site. $\Delta d_{\parallel}$ and
$\Delta d_{\rm z}$ represent, respectively, lateral and vertical
relaxations with respect to the bulk positions of the atoms
 indicated by the letters in parenthesis
(see Fig.~1).}
\end{table}

\newpage
\begin{table}
\begin{tabular}{l|c}
&  $(2 \times 1)$-O/Ru\,(0001)  \\
\hline
Structural parameters (\AA) & \hspace{0.8cm} O-Ru \hspace{0.5cm}
  $\Delta d_{z}$(A) \hspace{0.3cm} $\Delta d_{z}$(B)
 \hspace{0.4cm} $\Delta d_{z}$(C) \hspace{0.5cm} $\Delta d_{z}$(D)\\
\hline
LEED  & \hspace{0.90cm}  2.02 \hspace{1.0cm}
 $-$0.03 \hspace{1.04cm} 0.04 \hspace{0.4cm} $-$0.01 \hspace{1.0cm}
0.02 \\
   DFT-GGA &\hspace{0.9cm}   2.07 \hspace{1.0cm}
$-$0.06 \hspace{1.2cm}0.01 \hspace{0.5cm}$-$0.02\hspace{1.15cm} 0.00 \\
\hline
Structural parameters (\AA) & \hspace{0.5cm} $\Delta d_{\parallel}$(O)
 \hspace{0.5cm}
$\Delta d_{\parallel}$(A)
 \hspace{0.2cm} $\Delta d_{\parallel}$(B)
\hspace{0.3cm}  $\Delta d_{\parallel}$(C)
\hspace{0.3cm} $\Delta d_{\parallel}$(D)  \\
\hline
LEED  & \hspace{0.8cm}  $-$0.06 \hspace{1.1cm}$-$0.07
 \hspace{1.1cm}0.05 \hspace{1.0cm}0.05 \hspace{0.9cm} 0.04 \\
  DFT-GGA &\hspace{0.8cm} $-$0.03 \hspace{1.1cm}$-$0.02
\hspace{1.1cm}0.05 \hspace{1.0cm}0.01 \hspace{0.6cm} $-$0.02 \\
\end{tabular}
\caption{Structural parameters for $(2 \times 1)$-O/Ru\,(0001)
with O in the hcp-hollow site.
$\Delta d_{\parallel}$  and $\Delta d_{\rm z}$  represent,
 respectively, lateral and vertical
relaxations with respect to the bulk positions
of the atoms indicated by the letters in parenthesis
(see Fig.~2).}
\end{table}

\begin{table}
\begin{tabular}{l|l|l}
 & \multicolumn{2}{c}{$(1 \times 1)$-O/Ru\,(0001) }  \\
\hline
Structural parameters (\AA) &O-Ru \hspace{0.8cm}  $d_{\rm z,1}$
\hspace{0.9cm}
$d_{\rm z,2}$ \hspace{1.0cm} $d_{\rm z,3}$
\hspace{1.0cm}  $d_{\rm z,bulk}$ & E$_{b}$ (eV)\\
\hline
DFT-GGA (40 Ry) & 2.04 \hspace{1.0cm} 1.28 \hspace{0.9cm}
2.27 \hspace{0.9cm}   2.19 \hspace{0.9cm}   2.19& 4.84 \\
DFT-GGA (60 Ry) & 2.03 \hspace{1.0cm} 1.26 \hspace{0.9cm}
 2.25 \hspace{0.9cm}   2.17 \hspace{0.9cm}   2.19 & 4.87\\
\end{tabular}
\caption{Structural parameters for $(1 \times 1)$-O/Ru\,(0001)
with O in the hcp-hollow site.
O-Ru, $d_{\rm z}$, and E$_{b}$ represent, the O-Ru bond length,
 the interlayer spacings, and
binding energy, respectively.
}

\end{table}

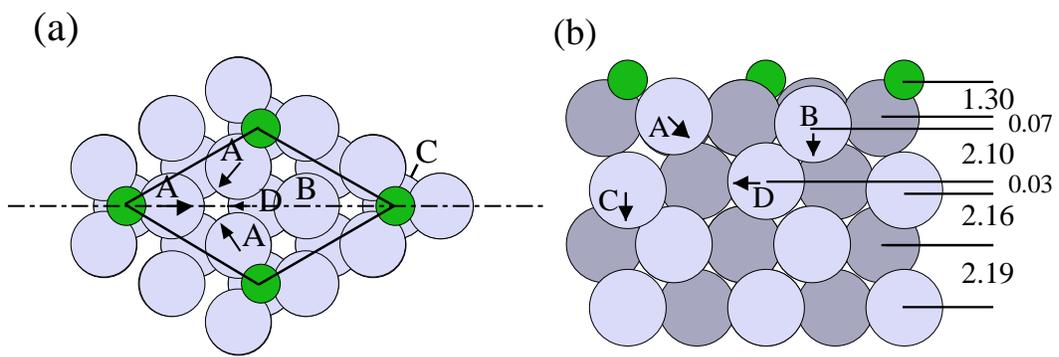
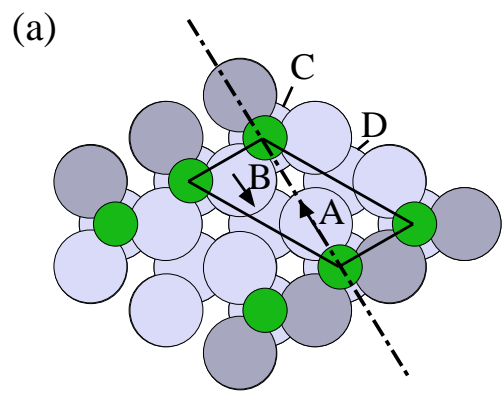
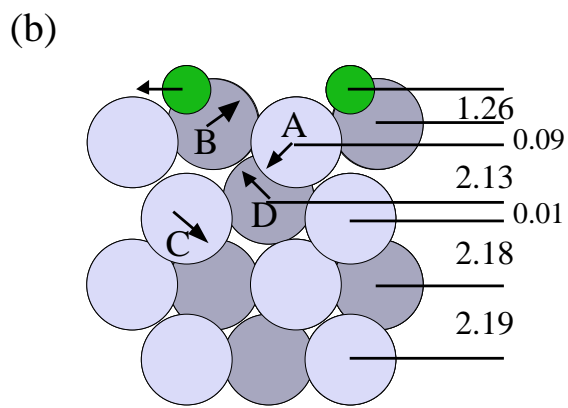
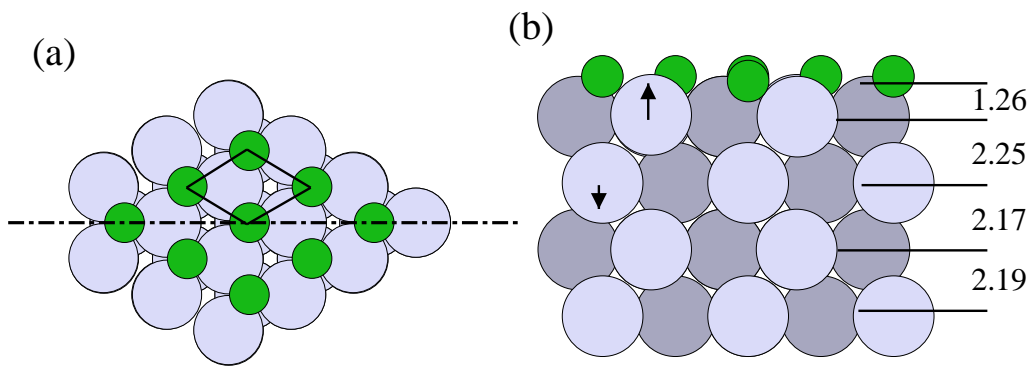
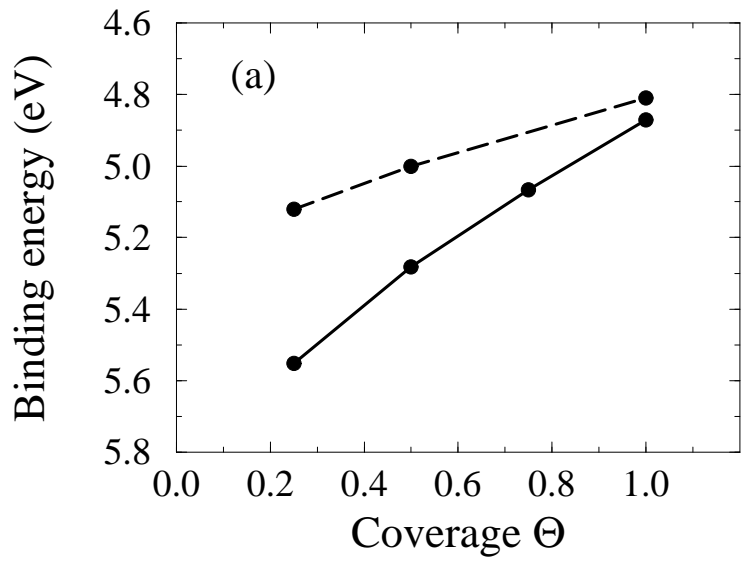
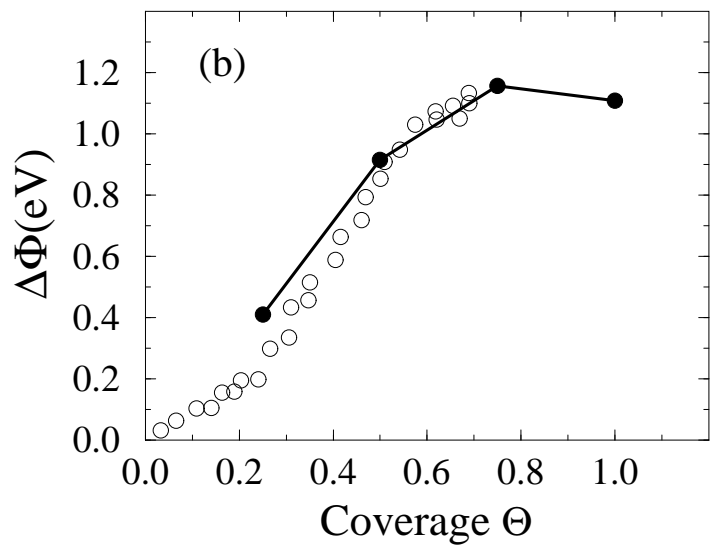
\end{document}